\documentstyle[12pt]{ioplppt}          
\begin{document}

\def\a{\alpha}
\def\bs{\bigskip}
\def\cl{\centerline}
\def\ra{\rightarrow}
\def\al{\alpha}
\def\be{\beta}
\def\ga{\gamma}
\def\de{\delta}
\def\la{\lambda}
\def\t{\tau}
\def\cl{{\cal L}}
\def\si{\sigma}

\def\both{\leftrightarrow\ }

\def\bege{\begin{equation}}
\def\ende{\end{equation}}

\title{Two-way traffic flow: exactly solvable model of traffic jam}

\author{H.-W. Lee\dag, V. Popkov\dag\ddag\ftnote{3}{To
whom correspondence should be addressed.} and D. Kim\dag\P }

\address{\dag Center for Theoretical Physics,
Seoul  National University,
	 Seoul 151-742, Korea}
\address{\P Physics Department,
Seoul  National University,
	 Seoul 151-742, Korea}
\address{\ddag\ Institute for Low Temperature Physics, Kharkov, Ukraine}

\maketitle

\begin{abstract}
We study  completely  asymmetric 2-channel exclusion processes in 1 dimension.
It describes a two-way traffic flow with cars moving  in  opposite directions.
The interchannel interaction makes  cars slow down in the vicinity of  
approaching cars in  other lane.
Particularly, we consider in detail the system with 
a finite density of cars on one lane and a single car on the other one.
  When the interchannel interaction
reaches a critical value, traffic jam occurs, which turns out 
to be of first order phase transition. 
We derive exact expressions for the average velocities, 
the current, the density 
profile and the $k$- point density correlation functions. We also obtain the
exact probability of two   cars in one lane being distance $R$ apart, provided 
there is a 
finite density of cars 
on the other lane, and show  the two cars form a weakly bound state in
 the jammed 
phase.

\end{abstract}

\section {Introduction}

Low - dimensional systems out of equilibrium attract much attention recently
\cite{Zia}. An important class of such models is the one-dimensional (1-D)
exclusion 
processes describing particles hopping independently with hard-core repulsion
along 1-D lattice. 
Such systems provide good description of growth processes, traffic flow and
queueing problems \cite{Nagel}, etc. (see  \cite{Zia} for the references up to 1995). 
The completely asymmetric exclusion process (ASEP) describing
 particles hopping only  to 
the right with equal rate 1 and hard core repulsion is perhaps 
the simplest and the best studied
one \cite{NATO},\cite{Derrida2}. In particular, for the periodic boundary 
condition, all configurations are equally likely in the steady state, and
  the average particle velocity in the infinite system  is
\bege
\langle v \rangle = 1 - n 
\label{velocity}
\ende 
$n$ being 
the density of particles. Janowsky and Lebowitz \cite{impurity} 
showed  that a fixed blockage in the system that reduces
the rate of hopping across it from 1 to $r<1$ can produce 
global effects. Namely, for each fixed density $n$, there is a range of
$0<r \leq r_0$ where the system segregates in high- and low- density regions
with sharp boundary, called shock, between them. Although some exact results were 
obtained \cite{impurity}, many quantities of interest e.g. steady state 
density profile, correlation functions, etc. were computed only numerically.

\bs

In the  present paper we derive all these quantities exactly, in closed form,
for a slightly different model, guided by modelling the two-way traffic
flow problem. Namely, there are two 1-D chains on a ring, $N$ sites 
each. One chain, or lane is occupied by  cars and another with trucks (we call 
them differently just for notational simplicity), 
hopping in opposite directions with rate $1$ and $\ga$, respectively.
 Effective rate of 
hopping of the car (truck) reduces to $1/\be (\ga /\be)$, 
when there is a truck (car) in front in another lane.  
 Physically, $1/\be$ is determined by narrowness of the road; it
describes how much car/truck slows down seeing another truck/car
approaching. ${1 \over \be}=0$ case corresponds to the road completely blocked. 

\bs

For the case of a single truck in  one lane and finite density of cars
$n$ in the other,
we expect the similar type of behaviour with the blockage case \cite{impurity}.
We expect the cars to pile up causing traffic jam, at a certain range of interlane
interaction parameter $\be$. So it is, as Monte-Carlo simulations unambigiously show. 
Then, to study closely the nature of traffic jam phase transition, we impose
the restriction that car and truck cannot occupy each 
neighbouring site $i$ simultaneously.
At this point, the model becomes exactly solvable by the matrix approach
of Derrida et. al. \cite{Derrida1,Derrida}. 
Using it, we compute the average velocities,
the density profile, $k$-point correlation functions exactly, for the finite chain
and in the thermodynamic limit. 
 Particularly, the traffic jam phase transition 
curve is given by simple formula 
$\be_{\rm crit}=  \ 1/n$, in the thermodynamic limit. 
 
Characterictics of the traffic jam phase transition are examined in detail.
 Finally, we consider the situation  with two trucks/finite density of
cars  in the system. We observe  that a weak bound state 
is formed between the two trucks.
\bs

Comparison of our results with the original two-way model, 
the ASEP with single fixed blockage \cite{impurity},
 and exact Bethe ansatz solution by Sch\"{u}tz \cite{Gunter}  for deterministic ASEP with
blockage concludes the paper.

\section{Formulation of the 2-way traffic flow model}

We consider the following hard core exclusion process:
there are two parallel 1-D chains on a ring, $N$ sites each,
the first  chain containing $M$ cars and the second one containing  $K$ trucks.
 Cars (trucks)  are hopping in opposite directions with rates $1$ ( $\ga$)
respectively. The state of the system is characterized by the set of
occupation numbers $\{ \t_i \}_{i=1}^N$ of the first lane 
 and    $\{ \si_i \}_{i=1}^N$ of the second lane. $\t_i =1$ if 
there is a car at site $i$ and  $\t_i =0$ if  empty, and the
same for the trucks,  $\si_i =1(0)$ if 
occupied (empty).  The system evolves under the following stochastic
dynamical rules:

At each infinitesimal time interval $dt$, one pair of adjacent sites $i,i+1$   
at any of the two chains is  selected at random for a possible
exchange of states. The possible exchange processes together
with their rates are listed below:

\noindent cars are hopping to the right
\bege
(\t_i,\t_{i+1}) = (1,0) \ra (0,1)\  \mbox{with rate} \  
\left\{
\begin{array}{ll}
1 & \mbox{if $\si_{i+1} =0$}  \\
{1 \over \be} & \mbox {if $\si_{i+1} =1$  (truck in front)} 
\end{array}
\right.
\label{car}
\ende
trucks are hopping to the left
\bege
(\si_i,\si_{i+1}) = (0,1) \ra (1,0)\  \mbox{with rate} \  
\left\{
\begin{array}{ll}
\ga & \mbox{if $\t_{i} =0$}  \\
{\ga \over \be} & \mbox {if $\t_{i} =1$  (car in front)} 
\end{array}
\right.
\label{tru}
\ende

\noindent One sequence of $2N$  (total number of sites in two chains) selections
constitutes
one time step (or one Monte Carlo step, see Figures 1,2).

The interlane interaction parameter $\be >1$ 
has a transparent physical meaning.
It describes how much a vehicle slows down seeing another vehicle 
approaching, which in turn depends on narrowness of the road. $\be =1$
(no slowing down) corresponds to a highway with divider, and $1/\be = 0$
corresponds to a narrow road completely blocked. 
If we let the system evolve for a long time, it reaches the steady state,
independent of its initial configuration.
 We shall be interested in the steady state
characteristics which depend only on macroscopic parameters
(number of cars, number of trucks and total number of sites) and the rates 
(\ref{car},\ref{tru}). 

We studied the most practical characteristics of the model, the 
average velocities of cars $\langle v_{\rm car}\rangle$ and
 trucks  $\langle v_{\rm truck}\rangle$
as a function of $(1 - 1/\be)$ by Monte-Carlo (MC) simulations.  The MC 
results for two different cases are shown in Figs. 1 and 2. Fig.1 corresponds
to the system with the density of cars and trucks being 0.3 and 0.4,
respectively, and shows monotonic decrease of both velocities as $\be$ 
increases. Fig.2 corresponds
to the system with  a single truck and many cars with density
$n=0.3$. In contrast to Fig. 1,
we see that   $\langle v_{\rm car}\rangle$ stays constant (equal to the 
average velocity (\ref{velocity}) in the 
noniteracting  system $v= (1-n)$, until the point $1 - {1 \over \be_{\rm c}} 
\approx 0.8$ is reached. For $\be > \be_{\rm c}$,  $\ \langle v_{\rm car}\rangle$ 
rapidly drops. Simulations show that for  $\be > \be_{\rm c}$, the
system segregates in two phases: the high density one in front
of the truck (traffic jam) and  the low density one behind the truck.
Piling up of cars in front of the truck accounts for the decrease of
average car velocity. 
Absence of sharp transition for a finite density of trucks (Fig.1)
compared to Fig.2 is due to the fact that
a finite number of trucks in infinite systems produces the macroscopic jammed 
phase, while a finite density of trucks produces only microscopic
jams which average out to give smooth behaviour.

The segregated or traffic jam phase is well known as a shock phase
or coexistence  phase  
in 1-D asymmetric exclusion processes (ASEP).   
Sch\"{u}tz \cite{Gunter} showed its
existence in an exactly solvable deterministic model with a fixed blockage, 
and found various shock characteristics rigorously.
As far as the probabilistic  ASEP are concerned,  Janowsky and Lebowitz 
\cite{impurity} showed the existence of segregated phase in a 
probabilistic ASEP with a fixed blockage. 
The latter model is not solvable, and 
most results  obtained in \cite{impurity} are therefore numerical.

The characteristics of the shock are believed to be quite universal, 
qualitatively independent on details of stochastic process. That is why
it is important to give exact solutions for some system with a shock. 
Here in this paper we propose a two-way traffic flow problem (slightly
modified, see below) with a single truck
 as an example of such a solvable system.
Roughly, the single truck plays the role of blockage and ${1 \over \be}$
plays the role of transmission coefficient $r$ in  \cite{impurity}. 
Following the traffic flow formulation,
we shall call the shock the `traffic jam' and the segregated or coexistence phase
the `traffic jam phase'. We shall find exactly the characteristics 
of the traffic jam and the traffic jam transition, including
 average velocities,
 density profiles, $k$-point correlation functions, for finite chains
and in the thermodynamic limit. To do so, one has to modify 
the original model to a solvable one.

\section {Modification of  
the original model to exactly solvable model}

We now  modify the two-way traffic problem slightly. Here, we
 forbid the car and the truck
to occupy the same site $i$ in the neighbouring chains simultaneously.
Then one can in fact describe the configuration by a single lane configuration 
$\{ \t_i \}_{i=1}^N;$  each 
site $i$ is either occupied by a car $\t_i=1$ or truck  $\t_i=2$ or empty  $\t_i=0$.
The allowed exchange processes is then modified from
Eqs.(\ref{car}) and (\ref{tru}) to:
\begin{eqnarray}
& (1, 0)  \ra (0, 1) \ \ \ &\mbox{with rate $1$} \nonumber \\
& (0, 2)  \ra (2, 0) \ \ \ &\mbox{with rate $\ga$}
\label{derrida} \\
& (1, 2)  \ra (2, 1) \ \ \ &\mbox{with rate ${1 \over \be}$}  \nonumber
\end{eqnarray}
Although the quantitative characteristics of the system do change after
this modification, the qualitative  characteristics do not (compare for instance
the graphs for average car velocities in Figs. 1 and 3).

\bs

The process (\ref{derrida}) 
is the two-species ASEP solvable by the approach of Derrida et al
\cite{Derrida}. 
The process (\ref{derrida}) and the one considered in \cite{Derrida}
differ by  replacement $2 \both 0$ (interchange of
trucks and empty spaces). 

The probability of a given steady state configuration
is shown in Ref.\cite{Derrida} to be (up to normalization) the trace of a product

\bege
w_{\rm conf}(\t_1 \t_2 \ldots \t_N) = Tr \left( X_1 \ X_2 \ldots X_N \right),
\label{weight}
\ende
where 
\bege
X_i = \ \ 
\left\{
\begin{array}{llll}
&D  &\mbox{if car at site $i$}   &\t_i=1 \\
&E  &\mbox{if truck at site $i$} &\t_i=2 \\
&A  &\mbox{if site $i$ is empty} &\t_i=0 
\end{array}
\right.
\nonumber
\ende
are noncommuting matrices satisfying the following algebra:
\begin{eqnarray}
&D E      &= D + E  \nonumber \\
&\be D A  &=  A  \label{algebra} \\
&\al A E      &= A; \ \ \ \ \ \al = \be \ga  \nonumber 
\end{eqnarray}
Knowing the probabilistic measure, we can find the various
averages of steady state.

\section {Average velocities}

We shall consider the system having $M$ number of cars and a single truck.

Analogously to  \cite{Derrida}, 
define $Y(N,M)$ as the probability of having the truck at site $N$, in a system with
$N$ sites and $M$ cars. Define then 
  $Y_D(N,M)$ as the probability of finding a car at site $N-1$, provided the truck 
occupies the position $N$. 
Then, the average velocities of the cars and the truck  are given
by:
\bege
\langle v_{\rm car}\rangle = {1 \over \be} 
{
Y_D(N,M) + (N-M-1) Y(N-1,M-1) 
\over
M \ Y(N,M)
}
\label{v_car}
\ende
\bege
\langle v_{\rm truck}\rangle = {1 \over \be} 
{
Y_D(N,M) + \al \left( Y(N,M)- Y_D(N,M) \right)
\over
Y(N,M)
}
\label{v_tru}
\ende
The quantities $Y(N,M)$ and  $Y_D(N,M)$ are computed in Appendix and found to
be\footnote{More precisely,  $Y(N,M)$ and  $Y_D(N,M)$ are probabilities
up to normalization factor which is equal for all the terms entering 
Eqs.(\ref{v_car}) and (\ref{v_tru}), see Appendix} 
\bege
Y_D(N,M) ={1 \over \al \be^M } \left( { - \al \over \be -1 } C_{N-2}^{M-1} + 
{  \al +\be -1 \over \be -1 } I(N,M) \right)
\label{Y_DNM}
\ende
\bege
Y(N,M) = Y_D(N,M) + {1 \over \al \be^M } 
 C_{N-2}^{M}
\label{YNM}
\ende
where 
\bege
I(N,M) = \sum_{k=1}^M \be^k \ C_{N-2-k}^{M-k}
\label{INM}
\ende
and $ C_{i}^{j}$ is the binomial coefficient.
In Figs. 3 and 4 we plotted 
$\langle v_{\rm car}\rangle$
and $\langle v_{\rm truck}\rangle $, respectively, computed from exact formula,
   as a function of $1 - 1/\be$, for three densities $n = 0.3$, $0.5$ and $0.7$
at $\ga = 1$,
in a system with $N$= 200 sites. Naturally the traffic jam transition
point decreases as average density $n$ increases. 
Behaviour of $\langle v_{\rm car}\rangle$ is similar to the 
one for the original two-way traffic model Fig.2.
However now we can evaluate exact thermodynamic limit $N,M \ra \infty$,
$\ n = M/N $ fixed, and find the exact transition
point.
 We used  the steepest descent method for 
computing the thermodynamic limits.   The average velocities in
the thermodynamic limit are given by (see Fig. 5):

\bege
\langle v_{\rm car}\rangle = 
\ \ \left\{
\begin{array}{lll}
&1-n  &\mbox{if $n \be \leq 1$ }  \\
&{1 \over \be} \ {1-n \over n}  &\mbox{if $n \be \geq 1$}
\end{array}
\right.
\label{v_car_therm}
\ende

\bege
\langle v_{\rm truck}\rangle = 
\ \ \left\{
\begin{array}{lll}
&{1 \over \be} \ 
{    \al(1-n)(1 - n \be) + n (\al+\be - n \be)
\over
  (1-n)(1 - n \be) + n (\al+\be - n \be) }
          &\mbox{if $n \be \leq 1$}  \\
&{1 \over \be}   &\mbox{if $n \be \geq 1$}
\end{array}
\right.
\label{v_tru_therm}
\ende

Thus the transition point to the jammed state
is given by a simple formula
\bege
n \be_{\rm crit} = 1
\label{be_c}
\ende 
Note that the transition point  does not depend on $\ga$ --- 
the free velocity of the truck. The average car velocity 
has a cusp at the transition point.  
The average velocity of cars before the transition  $n \be <1$ 
is equal to the one in a system without truck (\ref{velocity}).
For $n =1$,  $\ \ \ \ \langle v_{\rm car}\rangle \equiv 0$ independently of $\be$, 
because all sites are 
filled and cars cannot move. $1/\be=0$ is the case of complete blockage: 
both velocities are identically zero.
    
To examine closely the nature of the traffic jam transition, we find the exact 
 density profile and $k$-point correlation functions
 in the next two sections.

\section{The density profile}

In this section, we obtain the exact density profile
$\langle n(x)\rangle$ in a system with one truck and arbitrary $M$ number of cars,
in a chain of length $N$. 
We choose a reference frame in which truck is always at the position $N$.
It can be done because the weights of the steady state configurations
(\ref{weight}) depend only on the positions of the cars relative to 
the truck location, due to cyclic invariance of (\ref{weight}).
The average density $\langle n(x)\rangle$  at distance $x$ from the truck 
is equal to the probability of 
finding a car at site ${N-1-x}$;
\bege
\langle n(x)\rangle = {\sum_{\rm conf}\t_{N-1-x}\  w_{M}(\t_1 \t_2 \ldots \t_{N-1} \ 2)
\over \sum_{\rm conf}  w_{M}(\t_1 \t_2 \ldots \t_{N-1} \ 2) }
\label{n}
\ende
 Sums run over all possible 
configurations having  $M$ cars, $N-M-1$ empty spaces, with 
truck at the position $N$, so
\bege
 w_{M} ={\rm Tr} 
\left(
(\prod_{i=1}^{N-1} X_i) \ E   
 \right)
\nonumber  
\ende
where $X_i = D (A)$ if site $i$ is occupied by a car (empty).
The quantity (\ref{n}) is readily obtained using algebra (\ref{algebra})
and found to be

\def\Co{{\al \over \be -1}}
\def\CCo{{\al+ \be -1 \over \be -1}}
\begin{eqnarray}
& &\langle n(x)\rangle  = 
\{
 C_{N-3}^{M-1} - \Co  C_{N-3}^{M-2} +   
 \CCo [ I(N-1,M-1)+      \nonumber  \\
& & 
 \be^{x-1} (\be-1) I(N-x,M-x) \Theta(M \geq x)
]
\} / \left(Y(N,M) \al \be^M \right) 
\label{nfinite}
\end{eqnarray}
where 
$Y(N,M)$ and $I(N,M)$
 are given by Eqs. (\ref{YNM}) and (\ref{INM}), respectively, and 
\bege
  \Theta(y \geq x) =
\left\{ 
\begin{array}{ll} 
1, & \mbox{if $y \geq x$} \\
0, & \mbox{otherwise}
\end{array}
\right. 
\label{theta}
\ende
The expression (\ref{nfinite}) determines average density for any 
value of $\be$. Below we shall 
consider the cases before and after 
the traffic jam phase transition separately.

\subsection{ Low density phase; $n \be < 1$.}
Before the  transition the presence of truck affects the system only locally
as is seen from Fig. 6, where the density profile 
is shown for $N = 400$. The density, otherwise constant, locally
increases only in a close vicinity of truck. 

Evaluating the formula (\ref{nfinite}) for large $M,N \gg 1;$ 
with  
${M \over N} = n$ fixed, we find
\bege
\langle n(x)\rangle = n 
\left(
1 + {(\al+\be-1)(1-n) \over 1-n+ \al n} (n \be)^{x} 
\right) 
\label{n_before}
\ende 
One sees that the local density disturbance
decays exponentially at a finite length scale
\bege
 \xi = |\ln ( n \be)|^{-1}
\label{xi}
\ende
Therefore, the relative size of disturbed region
vanishes as ${1\over N}$. 
In principle, by common sense one would expect the existence of low 
density region right behind the truck. However it is absent in exact
solution, as seen from 
Fig. 6. We do not have simple explanation for this fact. 
Analogous behaviour was observed in exactly solvable deterministic
exclusion process with a fixed blockage \cite{Gunter}.

\subsection{Traffic jam phase;  $n \be > 1$}

In this region, cars pile up before the truck as seen from Fig.7.
Increase of interlane interaction $\be$ leads to increase of traffic
jam length 
\bege
l = {L_{\rm jam} \over N} \approx {n \be -1 \over \be -1} 
\label{l}
\ende
In thermodynamic limit $ M,N \ra \infty, {M \over N} = n$,
the latter formula becomes exact, and the density profile becomes 
a step function:
\bege
\langle n(x)\rangle = 
\left\{
\begin{array}{ll}
1 & {x \over N} \leq  {n \be -1 \over \be -1} \\
{1 \over \be} & \mbox{otherwise.} 
\end{array}
\right.
\label{ntherm}
\ende
Note that the density in the low density region is equal 
to the critical density $n_{\rm crit} = 1/\be$, independently
of the average density $n$. The same behaviour is observed in \cite{Gunter}.

In fact, in the jammed phase, the only way the truck can move is
by the process $(1,2) \ra (2,1)$ (because the contribution of the
processes  $(0,2) \ra (2,0)$ becomes exponentially small in the large $N$ limit).
 In the slow truck/ many cars problem
treated in  \cite{Xiamen}, the same is true 
when the rate for  $(2, 0 ) \ra (0, 2)$ process as denoted by $\al$
in  \cite{Xiamen}
is zero.  Thus our model in the jammed case is a special case of  
\cite{Xiamen}, up to exponentially small corrections.
 However only simple characteristics  
were investigated in \cite{Xiamen}; correlation functions, as well as 
large $N \gg 1$ limits were not studied. 

More precisely, for large $N,M  \gg 1$, we find up to corrections
of order $N^{-{1 \over 2}}$,
\bege
\langle n(x)\rangle = 1 -  {1 \over 2}(1-{1 \over \be}) 
\left\{
1 + \mbox{\rm erf}
\left(
{x- N l \over \Delta \  \sqrt{N} }
\right)
\right\} + O\left( N^{-{1 \over 2}} \right)
\label{n_erf}
\ende

\[
\fl \mbox{with } \ \ l =  {n \be -1 \over \be -1}, \ \  
\Delta = { \sqrt{2 \be (1-n)} \over \be -1 } \ 
\mbox{and erf}(y) = {2 \over \sqrt{\pi}} \int_0^y e^{-t^2} dt.
\]

\noindent
This shows that the shock interface extends over a region of width 
$\sqrt{N}$.
More careful considerations however show that the 
real shock interface is sharp
and extends over only 2 consecutive sites, and
the apparent width of 
 $\sqrt{N}$  is
due to shock position fluctuations ( The shock position 
fluctuations of order  $\sqrt{N}$ were also observed in the probabilistic
exclusion process with a fixed blockage \cite{impurity}).
Indeed, the discrete version of the density gradient correlation,
$  \langle \Delta n(x_1) \Delta n(x_2)\rangle$ where 
$ \Delta n(x) = n(x+1) - n(x)$
vanishes if $|x_1-x_2|>1$.
(It follows directly from Eqs. (\ref{split}) and (\ref{fj})). 
This shows that the jammed 
phase is indeed segregated into two macroscopic regions ---
 the low density one on the left
with $n_{\rm low} = {1 \over \be}$ and the high density one on the right
 $n_{\rm high} = 1.$  The fact that  $n_{\rm high} = 1$ 
is due to the nature of 
the process we consider (see (\ref{derrida})): once cars pile up before
the truck, the  car-truck exchange processes do not create empty spaces.

Finally, note that one can choose the difference between the 
average densities in the macroscopic regions in front and behind the truck
$ \delta n = n_{\rm high} - n_{\rm low} $
as an order parameter, characterizing the traffic jam phase transition.
With respect to this order parameter,
the transition to the jammed phase is of the first order,
as seen from Eq.(\ref{ntherm});
\bege
\delta n = 
\left\{
\begin{array}{ll}
0 & \mbox{if $n \be <1$}  \\
1-{1 \over \be} & \mbox{if $n \be  \geq 1$} 
\end{array}
\right.
\nonumber
\ende

\subsection {The hydrodynamic approach}
\label{hydrodynamic}

The thermodynamic limit results  Eq.(\ref{ntherm}) can as well be obtained 
from simple hydrodynamic arguments. Supposing that the 
segregated phase contains
two macroscopic regions of length $l N$ and $(1-l) N$, with average
car velocities in these regions $1-n_{\rm high}$ and  $1-n_{\rm low}$, 
respectively (see Eq.(\ref{velocity})), one can 
write down a set of equations.
First, from the car conservation,
\[n_{\rm low} (1-l) + n_{\rm high} l = n\]
and next, from the current conservation in the reference frame
 of the fixed truck,
\bege
j = n_{\rm low} (1-n_{\rm low} + \langle v_{\rm truck}\rangle) = 
n_{\rm high} (1-n_{\rm high} + \langle v_{\rm truck}\rangle)
\label{j}
\ende
finally from the definition of the average car velocity
\[ n  \langle v_{\rm car}\rangle = n_{\rm low} (1-l) (1-n_{\rm low}) + 
 n_{\rm high} l (1-n_{\rm high})
\]

\noindent
Substituting the values $\langle v_{\rm car}\rangle$ and $\langle v_{\rm truck}\rangle$ from Eqs. (\ref{v_car_therm}) and 
(\ref{v_tru_therm}), and solving the above system of three 
equations, we obtain exactly the result
Eq.(\ref{ntherm}).

The correctness of the hydrodynamic arguments in  the thermodynamic limit 
is due to the fact that indeed cars behave like an ideal
gas of interacting particles; correlations vanish in  the thermodynamic limit
as shown in the next section. 

One can easily compute the current flowing through the truck. 
The phase diagram ``current versus ${1 \over \be}$ ''
(narrowness of the road) is given in Fig. 8.
With fixed $\be$, the current $j$ increases with the density $n$, as
\begin{eqnarray}
&j = n ( \langle v_{\rm truck}\rangle+\langle v_{\rm car}\rangle) =   \nonumber \\
&n \left(
{1 \over \be} \ 
{    \al(1-n)(1 - n \be) + n (\al+\be - n \be)
\over
  (1-n)(1 - n \be) + n (\al+\be - n \be) } 
 + (1-n)
\right)
\end{eqnarray}
until the critical density 
\bege
n_{\rm crit} = {1 \over \be} 
\label{n_crit}
\ende
is reached. After that, in the jammed phase, the current stays constant
(see Eq.(\ref{j})),
\bege 
j_{\rm max} = n_{\rm high} (1-n_{\rm high} + \langle v_{\rm truck}\rangle) = 
{1 \over \be} 
\nonumber
\ende
for all densities $n_{\rm crit} \leq n \leq 1$.

Note that there is a single critical density value Eq.(\ref{n_crit}),
in constrast to the models with fixed blockage \cite{impurity,Gunter},
where two  critical densities exist, $\rho_{\rm crit}$ and 
$\tilde{\rho}_{\rm crit} = 1 - \rho_{\rm crit}$. The reason is 
the following; the models considered 
in \cite{impurity,Gunter} have the particle-hole
symmetry which is broken in the model we consider.

\section {The $k$-point correlation functions }

Here we obtain the  $k$-point equal time correlation functions  in
the  steady state, 
in exact and asymptotic forms, for a system with one truck and $M$
cars. Analogously to Eq.(\ref{n}), one defines

\bege
\fl \langle n(x_1) n(x_2) \ldots n(x_k)\rangle = 
{\sum_{\rm conf}\t_{p_1}\  \t_{p_2}\ \ldots \t_{p_k}\ 
 w_{M}(\t_1 \t_2 \ldots \t_{N-1} \ 2)
\over \sum_{\rm conf}  w_{M}(\t_1 \t_2 \ldots \t_{N-1} \ 2) }
\label{corr}
\ende
with $p_j = N-1-x_j.$

\noindent
Here we take $x_1 < x_2 < \ldots < x_k$\footnote{ Equality
of some argument values, say, $x_1 = x_2$ simply lowers the order of
the correlation function by $1$ as seen from Eq.(\ref{corr}); 
$ \langle n(x_1) n(x_1) n(x_3) \ldots n(x_k)\rangle = 
\langle n(x_1) n(x_3) \ldots n(x_k)\rangle$ }.
Sums run over all possible 
configurations having  $M$ cars, $N-M-1$ empty spaces, with 
the truck at the position $N$. 

Calculation of Eq.(\ref{corr}) leads to the following surprising result:
\bege
 \langle n(x_1) n(x_2) \ldots n(x_k)\rangle = \langle n(x_k)\rangle -
\sum_{j=2}^k f_j(x_{k+1-j})
\label{split}
\ende
The $k$-point correlation function in fact splits into a sum of $k$ terms,
each one depending on a single argument! The exact form of   $f_j(x)$
is given by
\def\Co{{\al \over \be -1}}
\def\CCo{{\al+ \be -1 \over \be -1}}
\begin{eqnarray}
&f_j(x) = [ C_{N-2-j}^{M+1-j}  + 
\{  -\Co  C_{N-j-2}^{M-j}   +  \CCo [(\be-1)  \nonumber       \\
&\fl \times J(N-j,M-j,x-1) \Theta(x \geq 2) + 
\be   C_{N-j-2}^{M-j}] \} \Theta(x \geq 1)  
]/ \left(Y(N,M) \al \be^M \right)  
\label{fj}
\end{eqnarray}
where $ \Theta (x \geq y)$ and $Y(N,M)$ are given by Eq.(\ref{theta}),
(\ref{YNM}), respectively, 
and  
\[ J(N,M,x) =  \sum_{i=1}^{{\rm min}\ (M,x)} \be^i \ C_{N-2-i}^{M-i}.\] 

We shall show that in the limit of large $N,M \gg 1$,  $f_j(x)$
is given by a remarkably simple formula
\bege
 f_j(x) = \kappa^{j-1}\left( 1 - \langle n(x)\rangle \right); \ \ 
\mbox{with } \ \kappa = \mbox{min} \left( n,{1 \over \be} \right)
\label{fj_approx} 
\ende
Indeed let us consider the connected two-point correlation function,
\begin{eqnarray}
& \langle n(x_1) n(x_2)\rangle_C =  \langle n(x_1) n(x_2)\rangle -  \langle n(x_1)\rangle\langle n(x_2)\rangle = 
\nonumber \\
& \langle n(x_2)\rangle (1 -\langle n(x_1)\rangle) - f_2(x_1)
\label{connected}
\end{eqnarray}
according to Eq.(\ref{split}).
As $ f_2(x_1)$ does not depend on $x_2$, one can choose
any convenient $x_2$. Take the point $x_2$ infinitely far apart from $x_1$;
$\ \ \ x_2 \gg x_1$, so that the correlations between them vanish 
$ \langle n(x_1) n(x_2)\rangle_C = 0$. Then 

a) before the transition $n \be  \leq 1$, we have for 
$N,M \gg 1$, using  Eq.(\ref{n_before}) and  (\ref{connected}),
\[f_2(x_1) = n (1 - \langle n(x_1)\rangle)\]

b)  after the transition $n \be  > 1$ , using  Eq.(\ref{n_erf})
and imposing in addition $x_2 \gg Nl$ we obtain
$\langle n(x_2)\rangle = {1 \over \be}$, and 
\[f_2(x_1) = {1 \over \be} (1 - \langle n(x_1)\rangle)\]

\noindent  
which proves the  formula Eq.(\ref{fj_approx}) for $j=2$. 
Recursively, one obtains the 
asymptotic behaviour of other functions $f_3(x), \ldots, f_k(x)$
from  (\ref{split}).  The formula (\ref{fj_approx}) can also
be obtained directly from (\ref{fj}).

\bs

Finally, the $k$-point correlation function, connected part, is given from
Eqs.(\ref{split})(\ref{fj_approx}) as
\begin{eqnarray}
& \langle n(x_1)  \ldots n(x_k)\rangle_C = \langle n(x_k)\rangle -
\sum_{j=2}^k \kappa^{j-1} 
\left( 1 - \langle n(x_{k+1-j})\rangle \right) - \nonumber \\
& \langle n(x_1)\rangle \ldots \langle n(x_k)\rangle, \ \ \ 
\mbox{where }  \kappa = \mbox{min} \left( n,{1 \over \be} \right)
\label{corr_final}
\end{eqnarray}
for $N,M \gg 1, \ \ x_1 < x_2 <\ldots <x_k \ \ $ and
$\langle n(x)\rangle$ is given by (\ref{n_before}) and (\ref{n_erf})
for  $n \be < 1$ and   $n \be > 1$ respectively.
Thus the $k$-point correlation function for  $N,M \gg 1$ is determined completely
by the 1-point correlation functions .

As an example, consider the two-point correlation function,
\[ \langle n(x_1) n(x_2)\rangle_C = (\langle n(x_2)\rangle - \kappa) (1-\langle n(x_1)\rangle) \]
Before the transition,   at the low density phase $n \be < 1$,
$\kappa = n$, $\langle n(x_2)\rangle$ is given by (\ref{n_before}), and
  $ \langle n(x_2)\rangle - \kappa \sim (n \be)^{x_2}.\ \  $
So the correlation function decays exponentially with a length scale
\[  \xi = |\ln ( n \be)|^{-1}\]
Thus, in the low density phase, 
the two-point correlation function is nonzero only in a close 
vicinity of truck $x_1< x_2 \sim \xi$. For $x_2 \gg \xi$, 
$\langle n(x_1) n(x_2)\rangle_C \equiv 0.$ Thus, in the whole region 
$\xi \ll x_2 <N$, cars do not feel any correlations between each other
and behave like an ideal gas of particles.

\bs

In the traffic jam phase, $n \be >1$, the two-point correlation function 
\bege
 \langle n(x_1) n(x_2)\rangle_C = (\langle n(x_2)\rangle - {1 \over \be}) (1-\langle n(x_1)\rangle) 
\nonumber
\ende
stays nonzero, only if both $x_1$ and $ x_2 \ ( x_1<x_2)$ are in the region  
$x_1,x_2 \in [Nl - \Delta \sqrt{N},Nl + \Delta \sqrt{N}]$
as is seen from (\ref{n_erf}). Otherwise, either 
$  \langle n(x_2)\rangle \approx {1 \over \be}, \ \mbox{or} \ 
\langle n(x_1)\rangle \approx 1$ and the correlation function
$ \langle n(x_1) n(x_2)\rangle_C$ vanishes.

Again, one can say that the jammed phase indeed has a phase separation:
a) solid-like phase with density $1.$ b) ideal gas phase 
(no correlations  between the particles-cars) of density ${1 \over \be}$.
This explains why the simple hydrodynamic approach (see section
\ref{hydrodynamic}) leads to the correct results in the thermodynamic 
limit.

\section{The bound state between two trucks}

Here we shall consider the system having two trucks,
and $M$ cars, and determine the probability  $\Omega(R)$
of two trucks being at distance $R$ apart.  This
probability is proportional to
\bege
\Omega(R) \sim \sum_{\rm conf} w_{M}(\t_1 \t_2
 \ldots \t_{N-R-2} \ 2 \ \t_{N-R}  \ldots \t_{N-1} \ 2  ),
\label{Omega}
\ende
with the sum running over all possible 
configurations having  $M$ cars and two  
trucks at  positions $(N-R-1)$ and $N$.
So $R=0$ corresponds to two trucks being next to each other.
Due to the periodic boundary condition,
$0 \leq R \leq {N-1 \over 2}$.

The exact expression for  $\Omega(R)$ at finite $N$ is unwieldy
and we shall not present it here
(Typical behaviour of  $\Omega(R)$ are shown in Fig.9 for
$N=200, n = 0.3$, and $\be = 3, 5$) 
 Instead we shall write down
its asymptotics in each phase.

i) $n \be <1:$
\noindent In the thermodynamic limit,
 $\Omega(R)$ reduces to the following;
\bege
\Omega(R) \sim
1 + {n (1-n) (\al+\be-1) (\al-1) \over (1-n+ \al n)^2} (n \be)^{R} 
\label{Omega_before}
\ende 
It is maximal for $R=0$ and decays exponentially with the same length
scale $\xi = |\ln ( n \be)|^{-1}$, as before.
 As $\xi$ does not depend on $N$, the fraction 
of space with nonzero correlations between the trucks vanishes 
as $1/N$.  Thus two trucks are asymptotically free. The 
asymptotic freedom of trucks accounts for the fact, that the phase
transition to the jammed phase takes place at the same critical density,
$n_{\rm crit} \be = 1$. These arguments can be extended
to any finite number of trucks in the infinite system $N \ra \infty$.

ii) 
\noindent
Jammed phase $n \be \geq 1$: In the thermodynamic limit, 
 we find the following result for   $\Omega(R)$ 
from the exact formula:

$\Omega(R)$ linearly drops with the distance $R$ in the region
\[0 \leq R \leq N r_0, \ \ r_0 = {\rm min}(l,1-l) <{1 \over 2}, \ \ 
 \ \  l =  {n \be -1 \over \be -1};
\]
and then stays constant $\Omega(R) \equiv const$, 
$\ \  N r_0 \leq R \leq {N-1 \over 2} $, see Fig. 9(b). 
The relative ratios are 
\bege
\mbox{for} \  r_0 = l , \ \  \ \ {\Omega(N r_0) \over \Omega(0)} =
 \ \ 1 - {(\al-1) (\be-1) \over \al \be} 
\label{l<1/2}
\ende
\bege
\mbox{for} \  r_0 = 1-l , \ \ \ \ {\Omega(N r_0) \over \Omega(0)} =
 \ \ 1 - {(\al-1) (\be-1) \over \al \be} {r_0 \over 1-r_0 }
\label{l>1/2}
\ende
One can interprete this as the two trucks
 forming a weak bound state in the traffic jam phase.
The probability $\Omega(R)$ was studied also in paper \cite{Derrida},
for a system with two second class/many first class particles,
both hopping in the same direction, where it shows a power law
 decay $\Omega(R) \sim R^{-3/2}$ 
in the uniform background of the first class particles.
This is in marked contrast to our result Eq.(\ref{Omega_before}) 
showing exponential decay in the uniform low density phase.
Note however that  the asymptotic Eq.(\ref{Omega_before}) was obtained 
in supposition $\al \neq 1,\  \be \neq 1$, while  the results
of the paper  \cite{Derrida} are derived for  $\al = 1, \be = 1$
case.

It is interesting to analyze the average distance between two trucks.
Consider the case $n <0.5$ first. Analysis of (\ref{Omega_before}) and
(\ref{l<1/2}) shows
that in the thermodynamic limit the relative distance between the trucks
${<r> \over N } = 1/4$ in the low density phase $n \be <1$ and
then drops monotonically as function of $\be$ from ${<r> \over N}  = 1/4$ to 
${<r> \over N}  = n/3$ at $\be = \infty$ (complete blockage). 
At the same time the  length $l = (n \be-1)/(\be -1)$ increases
from $l=0$ at the transition point $\be_{\rm crit} = 1/n$ to $l=n$ at  $\be = \infty$.
We recall the reader that 
$ l =  {n \be -1 \over \be -1}$ is the total length of traffic jam in a 
system with 1 truck in the thermodynamic limit, see (\ref{l}).
One can argue in a different way that the  total length of
the  traffic jam is independent on the number of trucks, as long
as it remains {\it finite}. (e.g. by using the hydrodynamic approach,
see subsection \ref{hydrodynamic} ). 
That means in the early stage of traffic jam phase (small $\be$) 
$ {<r> \over N} > l $ and two separate 
traffic jams in front of the two trucks are formed, 
of lengths $l_1$ and $l_2$, $\ l_1+l_2 = l$ 
separated by the low density regions. As $\be$ increases, eventually
$ {<r> \over N} < l $, and the two 
 separate jams merge into a single one of  length $l$.


\section{Summary}

We have formulated the two-lane traffic model and showed that it has the transition
 from the low density phase to the 
segregated (traffic jam) phase. Modifying the model 
to an exactly solvable one, we studied in detail the
 characteristics
of the solvable model, which we believe to be qualitatively correct 
for the original one. The solvable model in the 
traffic jam phase in  $N \gg 1$
limit is a special 
case of the 2-species model considered by Derrida \cite{Xiamen}, see discussion 
after  Eq.(\ref{ntherm}). However our angle of view is different 
and most results obtained in sections 4-7 are new. We have obtained
 exact expressions of
 the current $j$, the average density profile, and the
 $k$-point correlation
functions, for finite chain, and in the large $N$ limit, for a single truck.
We have also studied the two trucks case and observed that a weakly bound state
is formed between them in the traffic jam phase. 
Generally, the truck slowing down the car movement, can be thought of as
a sort of 
moving blockage. Qualitatively our results for the density profile,
the current $j$ phase diagram, two-point correlation functions agree with those  
obtained in \cite{Gunter} and in part with those in \cite{impurity}, describing
a fixed blockage. However the last 
two systems possess the 
particle-hole symmetry and therefore the uniform high density phase, 
related to the 
low density one by this symmetry. In our case, the particle-hole symmetry
is broken for both our original two-lane model
and the modified solvable one. This accounts for the absence of the
 uniform high density phase in our model, see phase diagram Fig. 8.

\section*{ Acknowledgements}

V.P.  thanks B. Derrida for stimulating discussions and D. Dhar for  
references.   
 This work was supported
 by the Korea Science and Engineering
Foundation through the SRC program, and 
in part by the INTAS grant 93 -- 1324, 93 -- 0633,

\appendix

\section{Computation of $Y(N,M)$ and  $Y_D(N,M)$ }

The probability of finding the truck at the site $N$
is given up to normalization by 
\bege
Y(N,M) = 
\sum_{\rm conf}  w_{M}(\t_1 \t_2 \ldots \t_{N-1} \ 2)
\label{sum}
\ende
where this sum is over all possible 
configurations having  $M$ cars, $N-M-1$ empty spaces, with 
the truck at the position $N$. 
Split the above sum in two terms as
\bege
Y(N,M) = 
\sum_{\rm conf}  w_{M}^D(\t_1 \t_2 \ldots \t_{N-2} \ 1 \ 2) +
\sum_{\rm conf}  w_{M}^A(\t_1 \t_2 \ldots \t_{N-2} \ 0 \ 2)
\label{sum12}
\ende  
The first ( second ) term corresponds to a car (empty space)
being at site $(N-1)$. 
The second term can be written as
\bege
 w_{M}^A(\t_1 \t_2 \ldots \t_{N-2} \ 0 \ 2) = {\rm Tr}(C A E) = 
{1 \over \al}   {\rm Tr}(C A)
\label{wm1}
\ende
using the algebra (\ref{algebra}).
Here,
$C$ corresponds to arbitrary configuration of length $N-2$ 
having $M$ cars and $Q = N-M-2$ empty spaces. Generally,

\[CA = D^{m_1} A^{q_1} D^{m_2} A^{q_2} \ldots  D^{m_k} A^{q_k}\]
with $m_1+m_2 + \ldots + m_k = M$, 
 $q_1+q_2 + \ldots + q_k = Q$,
$q_k \geq 1$.

According to  (\ref{algebra}), 
\bege
{\rm Tr} (CA) = {\rm Tr} \left(
{1 \over \be^{m_1+m_2 + \cdots + m_k}} A^{q_1+q_2 + \ldots + q_k}
\right)
= {1 \over \be^{M} }  {\rm Tr} \left( A^Q \right)
\nonumber
\ende
and 
\bege
 w_{M}^A (\t_1 \t_2 \ldots \t_{N-2} \ 0 \ 2) = { 1 \over \al \be^M} 
{\rm Tr} \left( A^Q \right) 
\label{wma}
\ende
The first term in (\ref{sum12}) for some specific configuration is
\bege
w_{M}^D(\t_1 \t_2 \ldots \t_{N-2} \ 1 \ 2) =
{\rm Tr} \left(
 D^{m_1} A^{q_1} D^{m_2} A^{q_2} \ldots  D^{m_{k-1}} A^{q_{k-1}}
 D^{m_k} E
\right)
\label{wm2}
\ende
where  $m_1+m_2 + \ldots + m_k = M$,  $q_1+q_2 + \ldots + q_{k-1} = Q$,
$m_k \geq 1$. 
We have the following recursive relation: 
\bege
f_m = A D^m E = A  D^{m-1} (DE) =  A  D^{m-1}(D+E) =  A D^m + f_{m-1} 
\nonumber
\ende
Using the last expression recursively, one finds
\bege
f_m = A \sum_{i=1}^m  D^i + f_{0}; \  f_{0} = AE = {1 \over \al} A; 
\nonumber
\ende
Substituting the value of $f_m$ into (\ref{wm2}),
using  (\ref{algebra}), we obtain 
\bege
w_M^D  = {1 \over \al \be^M} \ga_{m_k} {\rm Tr} \left(A^Q \right)
\label{wmd}
\ende
where 
\bege
 \ga_{m} = \al \sum_{i=1}^m \be^{m-i} + \be^m = - {\al \over \be -1} +
 {\al + \be -1 \over \be -1} \be^m
\label{gamma}
\ende

The same common factor $ {\rm Tr} \left(A^Q \right)$ cancels from 
all formula for averages starting from (\ref{v_car},\ref{v_tru}) and
on, as it enters to both the denominator and numerator. 
Below we shall set $ {\rm Tr} \left(A^Q \right) = 1$ for 
simplicity. 

Using (\ref{wma}) and  \ref{wmd}) and some combinatorics to
count the number of configurations, the sum (\ref{sum12}) then reads

\bege
Y(N,M) =  { 1 \over \al \be^M} \sum_{m=1}^M \ga_m C_{N-m-2}^{M-m} + 
 { 1 \over \al \be^M} C_{N-2}^M
\nonumber
\ende
Finally, substituting (\ref{gamma}) and performing the summation  
${\displaystyle
\sum_{m=1}^M } C_{N-m-2}^{M-m} = C_{N-2}^{M-1}
$,
one obtains Eqs.(\ref{YNM}) and (\ref{Y_DNM}).

\Figures

Fig. 1. MC simulation result of the two-lane
traffic flow model.  Shown are the  ``average velocities
versus interlane interaction $r = 1 - {1 \over \be}$'' for 
 $N=200$, $\ n_{cars} = 0.3, n_{trucks} = 0.4$ and $\ \ga =1$
\footnote{$\ga =1$, unless otherwise stated}. 
Initial configuration of the system is random. We
 equilibriate the system for  
2000 MC step intervals, and collect data 
at 2000 MC step intervals and average over 100 different
histories.

\bs
\noindent
Fig. 2. The same as Fig. 1 but for 
$\ N_{cars} = M = 60$ and $ N_{trucks} = 1$.
The point $r_c \approx 0.8$ is the 
approximate traffic jam transition point.

\bs

\noindent
Fig. 3. Average velocities of the  cars  computed from
Eq. (\ref{v_car}) in a system of 200 sites, for 
different densities $n=0.3, \ \ 0.5, \ \ 0.7 $.

\bs

\noindent
Fig. 4.  The same as Fig. 3 
 for the the truck velocities Eq. (\ref{v_tru}).

\bs

\noindent
Fig. 5. Exact car and truck velocities in the thermodynamic limit
     Eqs.(\ref{v_car_therm}) and (\ref{v_tru_therm}), for $n = 0.3$. 
     $\langle v_{\rm truck}\rangle$ is given for $\ga = 1$ and 
     $\ga = 1.5$ (thin line).

\bs

\noindent
Fig. 6. Density profile before traffic jam, in a chain of 400 sites
computed from formula (\ref{nfinite}). $n = {M \over N } = 0.3, \ \ 
\be = 3 \ \ (\be_{\rm crit} = 3.3333 \ldots) $. The single truck 
is located at right end.

\bs

\noindent
Fig. 7.  Density profile in the  traffic jam phase, in a chain of 400 sites
computed from formula (\ref{nfinite}) for  $n = {M \over N } = 0.3, $.
and for  $\be = 5$(a) and $\be = 15(b)$. 
 $ (\be_{\rm crit} = 3.3333 \ldots) $.
The total length of traffic jam is 
$L_{\rm jam} \approx  {n \be -1 \over \be -1} N$. Asymptotic values of
densities in a high (low) density regions are $n_{\rm high} = 1, \ \ 
 n_{\rm low} = {1 \over \be}$.   Truck 
is located at the right end.

\bs

\noindent
Fig. 8.  Phase diagram of the solvable model in 
''current versus ${1 \over \be}$ '' plane. The curve
$j = {1 \over \be}$ corresponds to jammed phase
 $n_{\rm low}=  {1 \over \be};\ \ n_{\rm high}=  1$.
The area below corresponds to uniform low density phase 
$n_{\rm low}= n_{\rm high} < n_{\rm crit}$. There is no uniform
high density phase like in \cite{impurity,Gunter} 
because the particle-hole symmetry is broken.

\bs

\noindent
Fig. 9. Probability of finding two trucks at a distance $R$ apart,
before (a) and after (b) the phase transition, computed from
the exact formula,
for $N= 200, \ \  n =0.3$. The thin line in (b) corresponds to the 
thermodynamic limit. The value of $\be$ is $3$ and $5$ for
(a) and (b) respectively.

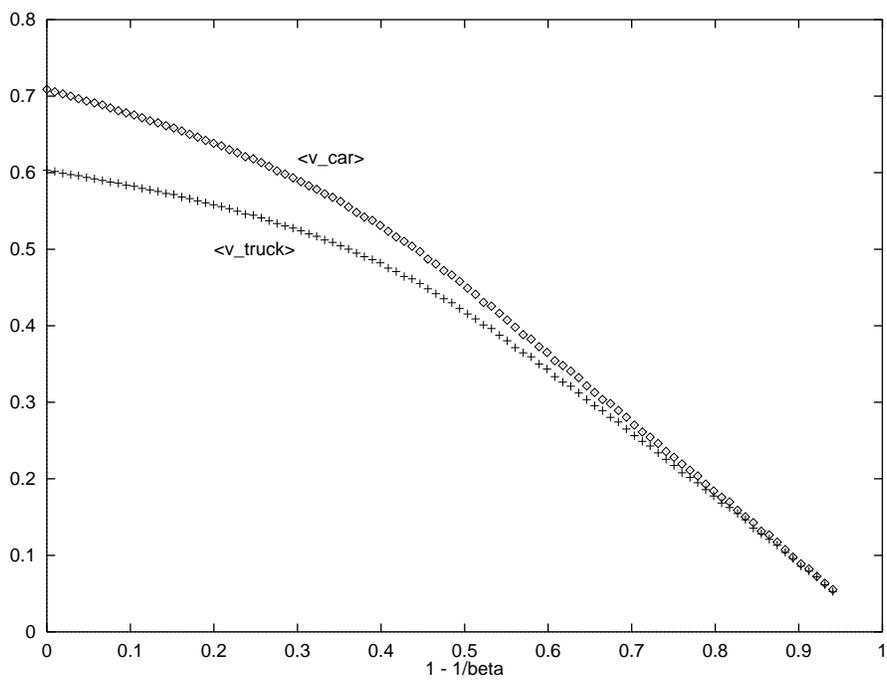
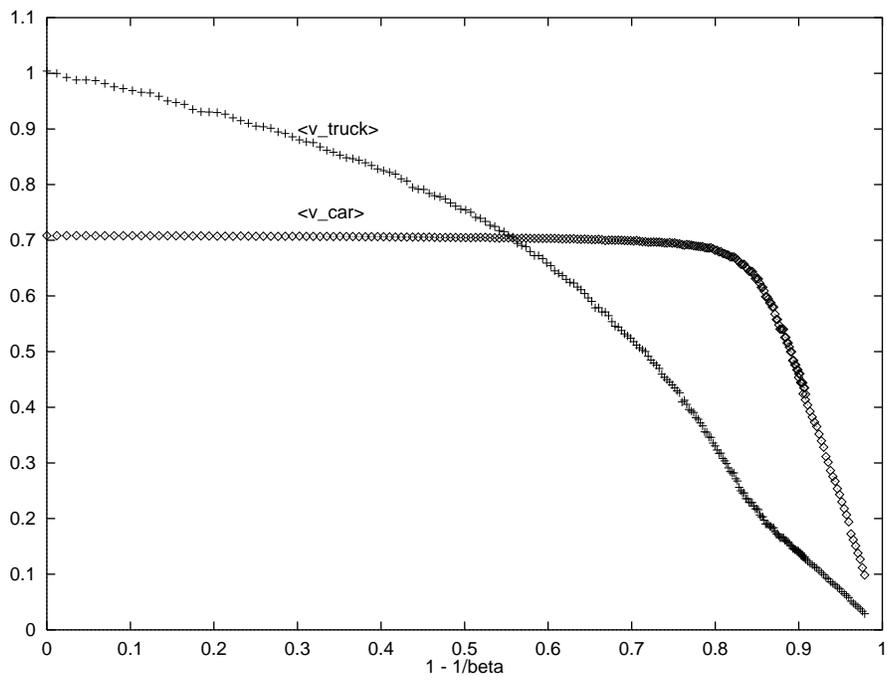
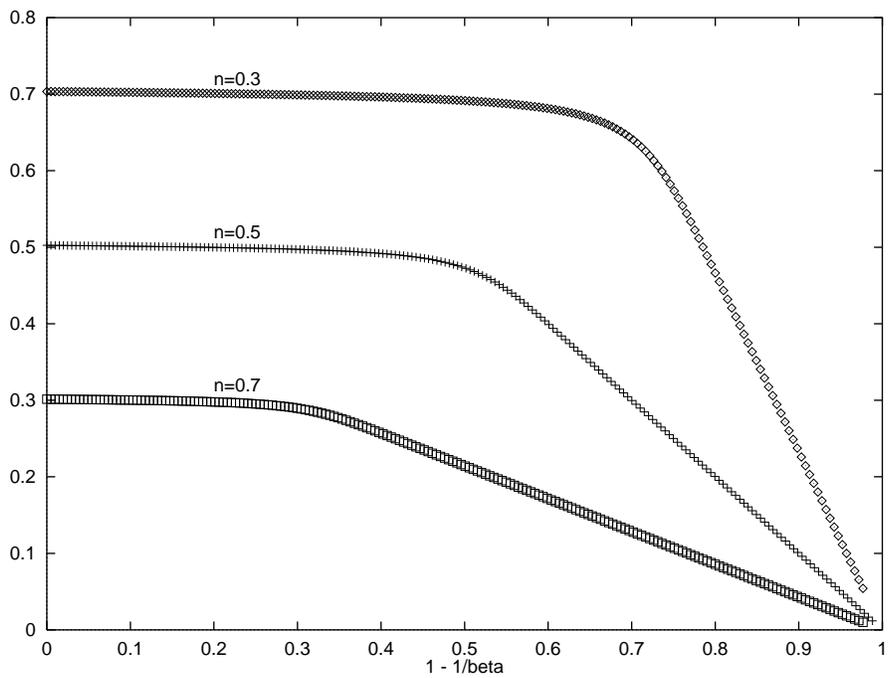
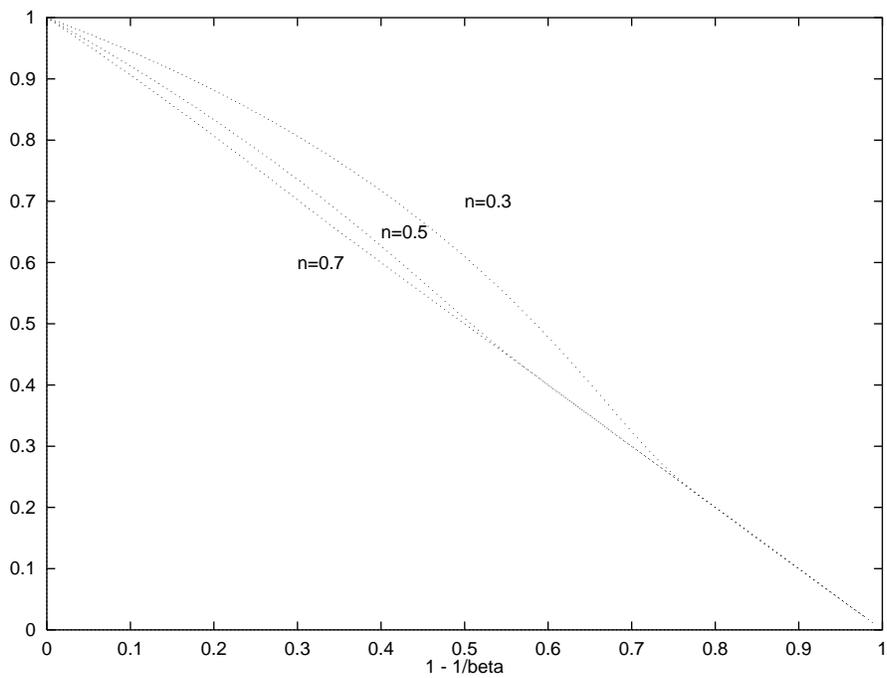
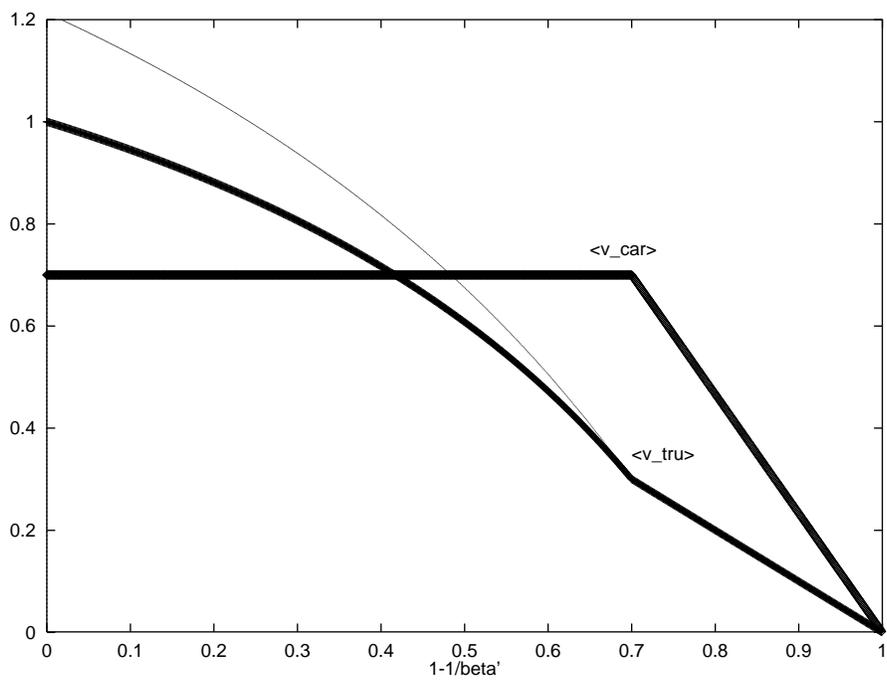
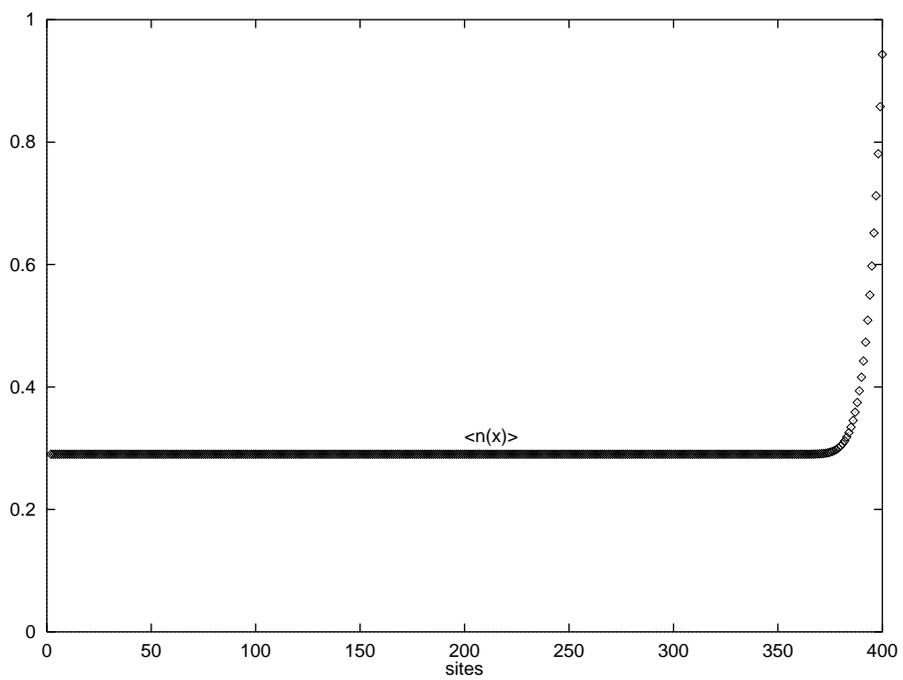
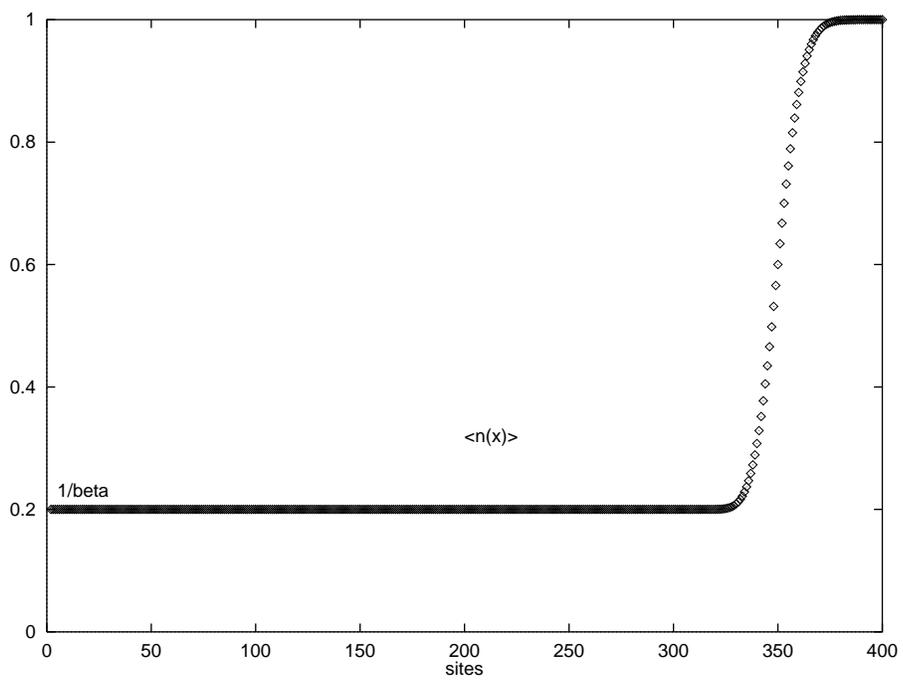
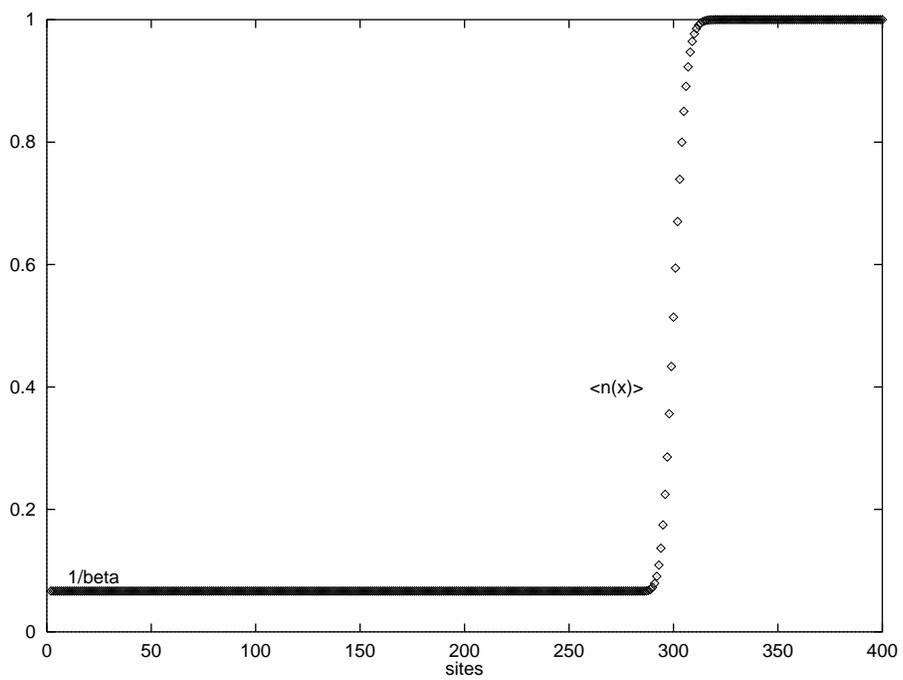
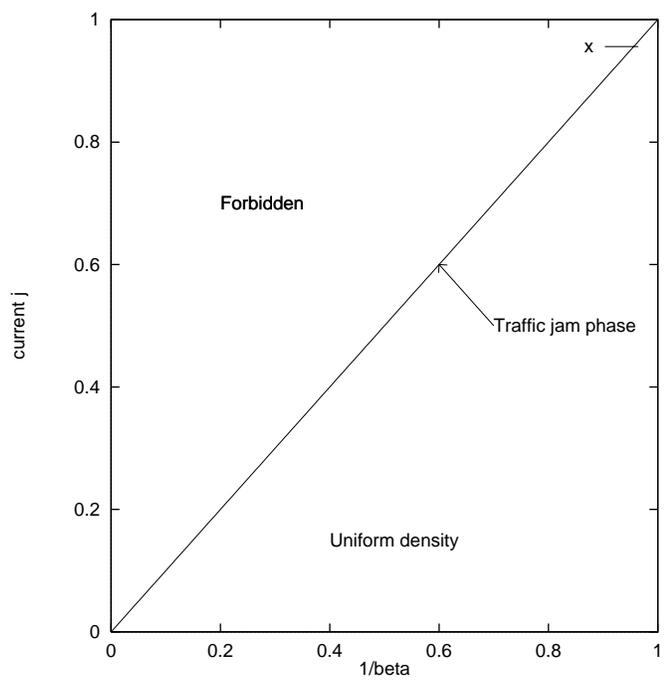
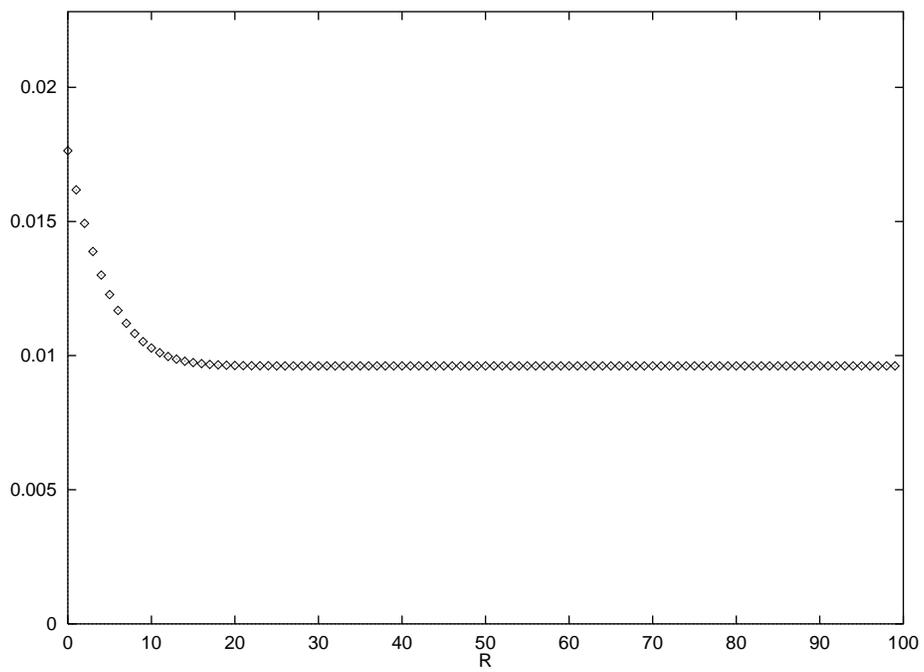
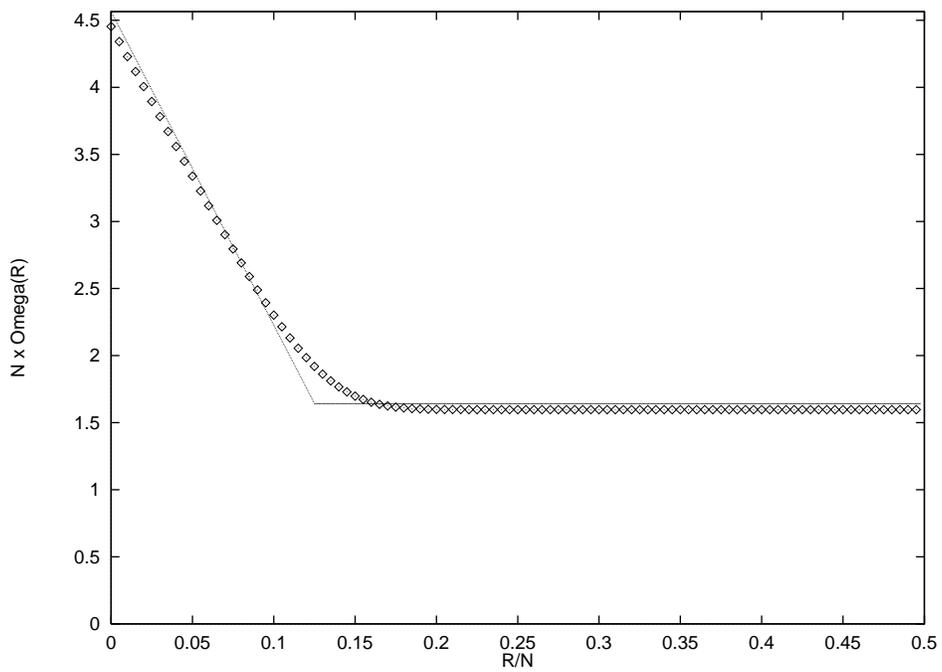
\end{document}